\documentclass[12pt]{iopart}
\usepackage{iopams}  
\usepackage{graphicx}
\usepackage[utf8]{inputenc}
\usepackage{tikz}
\usepackage[colorinlistoftodos]{todonotes}

\begin{document}

\title{Measuring light with light dependent resistors: an easy approach for optics experiments}

\author{F. Marinho$^{1*}$, C.M. Carvalho$^1$, F.R. Apolin\'ario$^1$, L. Paulucci$^2$}
\address{$^1$ Universidade Federal de São Carlos, Rodovia Anhanguera, km 174, 13604-900, Araras, SP, Brazil}
\ead{fmarinho@ufscar.br}
\address{$^2$ Universidade Federal do ABC, Av. dos Estados, 5001, 09210-170, Santo André, SP, Brazil}

\date{\today}

\begin{abstract}
 We entertain the use of light dependent resistors as a viable option as measuring sensors in optics laboratory experiments or classroom demonstrations. The main advantages of theses devices are essentially very low cost, easy handling and commercial availability which can make them interesting for instructors with limited resources. Simple calibration procedures were developed indicating a precision of $\sim 5\% $ for illuminance measurements. Optical experiments were carried out as proof of feasibility for measurements of reflected and transmitted light and its quality results are presented. In particular, the sensor measurements allowed to verify the angular distribution of a Lambertian reflective material, to observe transmitted and reflected specular light on a glass slab as function of the incoming angle of a light beam, and to estimate glass refractive index with values averaging $1.51\pm0.06$ in satisfactory agreement with the expected 1.52 value. 
\end{abstract}

%

\section{Introduction}

Light dependent resistors (LDRs) are widely used as devices to automatically control the functioning of our day-to-day streetlights, nightlights, alarms, among other applications. Although very cheap and abundant in the electronics component market it has not been much explored or at least documented on optics related learning activities. The characteristics of these sensors indicate that direct measurements of irradiance or illuminance (depending on the calibration equipment available) on simple made experiments are feasible. Its basic operation does not require electronics knowledge and data quality should allow comparison between theoretical expectations and experimental data.  

\begin{figure}[h!]
\centering
\includegraphics[scale=0.4]{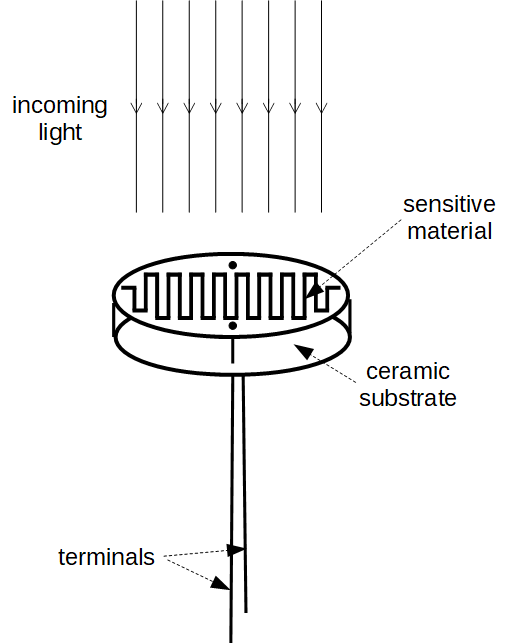}
\caption{Schematics of a light dependent resistor.}
\label{fig:ldr}
\end{figure}

The LDRs are passive devices that usually employ materials that exhibit detectable photoconductivity \cite{ldrdatasheet}. That is the ability to change their conductivity as the incident irradiance on the material varies. Semiconductors are often used as photoconductive materials which are deposited in a corrugated thin strip arrangement on top of a ceramic substrate. The strip extremities are attached to metal contacts leading to the pin terminals for connection. A transparent cover is usually placed on top of the surface for protection. Figure \ref{fig:ldr} shows a LDR schematics with its main components. When light is shed on the device excitation rate is such that the number of free charge carriers is increased and so is the electrical conductivity. Therefore, this relationship between incoming irradiance and conductivity (or resistivity) can be determined and calibrated such that these sensors can be used as measurement tool for didactic optics experiments.  

LDRs offer great advantage for being passive elements, thus being easy to use and measurements can be made directly with an ohmmeter only. No further connections or applied voltages are required. Acquisition rate is limited by the characteristic decay time of photo-electrons in the material. This time can be as high as tenths of a second which means LDRs should be better suited for slow changing irradiance situations and mechanically static setups.
Photodiodes and phototransistors can also be used as devices to measure irradiance, as done, for instance by Benenson \cite{benenson} among many examples. But for these, it is necessary to assemble an electric circuit with external power supply in order to measure an output voltage with incident light. 

Some applied uses of LDRs related to science and engineering instruction have been reported and are briefly described below. Rivera-Ortega and collaborators \cite{fibraotica} present an interesting work using a laser diode to convert a 4-bit image into pulses that were transmitted by an optical fiber, received by a LDR, and recorded by an Arduino board. The reading of the variable electric voltage on the LDR allowed the reconstruction of the image with good quality. Gutierre et al. \cite{rbef} performed the calibration of a LDR as a light sensor using a laser and polarizers and then used it to obtain the Gaussian light profile of a HeNe polarized laser beam with satisfactory results. Nevertheless, no further applications are discussed. A LDR was also employed as part of a simple spectrophotometer by Tavener\&Thomas-Oates \cite{tavener} and Forbes\&N\"othling \cite{forbes}, in the context of teaching chemistry to undergraduate students. Although not related to teaching optics, LDRs have also been used in other classroom situations, as in \cite{arduino}, where the voltage readings on the LDR terminals were used to decipher Morse code messages created by using a laser pointer. 

In all the aforementioned applications found so far, LDR sensors were used as parts of more sophisticated apparatuses with components that in some situations may not be readily available or might render many replicates in the laboratory little affordable. We propose and evaluate the use of LDRs for performing simple experiments to study typical optical phenomena measuring illuminance and its dependence with variables such as incidence and reflection angles. A calibration procedure to evaluate a LDR using low cost materials is described in section \ref{calib} and other experimental setups follow in section \ref{exps}. Discussion on the results, feasibility and possible usage in laboratory activities and classroom demonstrations are made in section \ref{concl}.

\section{Calibration}\label{calib}

A typical procedure to evaluate optical characteristics of a sample generally requires the use of a light source and a suitable detector. For our calibration studies we used a lamp placed at different distances to understand how the LDR sensor's measured resistance varies with illuminance. A commercial lux meter was used as calibrating device to provide measurements simultaneous to the LDR resistance readings. This way a correlation between the two quantities can be established. Because the placing distances used are significantly larger than the lamp dimensions the illuminance can be sufficiently varied and controlled by simply moving the source towards the detectors direction in small steps. This also allowed to consider light emission from the used source as isotropic.

\begin{figure}[ht!]
\centering
\includegraphics[scale=0.5]{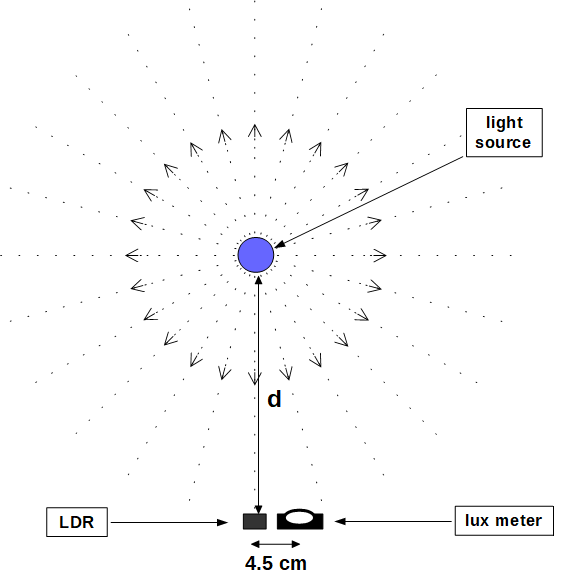}
\caption{Calibration setup. The lamp illuminates both the LDR and lux meter sensors, placed at a distance $d$ from it.}
\label{fig:calsetup}
\end{figure}

Figure \ref{fig:calsetup} shows a representation of the calibration setup. The source is placed on top, the lux meter and a $\rm 0.4 \times 0.5~cm^2$ LDR close at the bottom. A 6 Watt LED round red lamp was used as an isotropic light source. The red lamp was employed to verify the LDR response around the wavelength region close to the typical red laser pointer wavelength since these are the cheaper and easiest to find specular light beam sources.

A lux meter has a typical cost of about 10 to 30 US dollars. A smart phone light sensor can be used as an alternative calibrating device along with specific software (e.g. Lux Meter, Physics Toolbox, Lux Light Meter Pro, Lux Light Meter FREE) freely available for Android or IOS platforms.

Although the lux meter provided measurements of illuminance, i.e. the luminous flux per unit area as seen by the average human eye, we used a monochromatic light source such that the ratio between the actual irradiance arriving in the sensor and the illuminance is always given by a fixed conversion factor. 
Mathematically:
\begin{equation}\label{ill_vs_irr}
    I = \eta\int_{0}^{\infty} E_{e,\lambda}V(\lambda) d\lambda,
\end{equation}
where  $I$ is the illuminance, $\eta$ is a conversion constant, $E_{e,\lambda}$ is the spectral irradiance, $V(\lambda)$ is the luminous efficiency of the eye and $\lambda$ is the light wavelength. 

In our case:
\begin{equation}
E_{e,\lambda} \propto P\delta(\lambda-\lambda_{source}),
\end{equation}
where $P$ is the source power and $\delta$ is the Dirac delta function centered at the source wavelength, so that the integral becomes:
\begin{equation}
E_{v} \propto P.
\end{equation}

This is only a reasonable statement when employing monochromatic light sources.

A distance scan was performed and illuminances and resistances were recorded. Room illumination was much reduced for data acquisition. Readings compatible with $\rm 0.0~lx$ were obtained from the lux meter with the light source off corresponding to a resistance reading above $\rm 20~M\Omega$ for the LDR, in agreement to specifications in light absence.

\begin{figure}[ht!]
\centering
\includegraphics[scale=0.4]{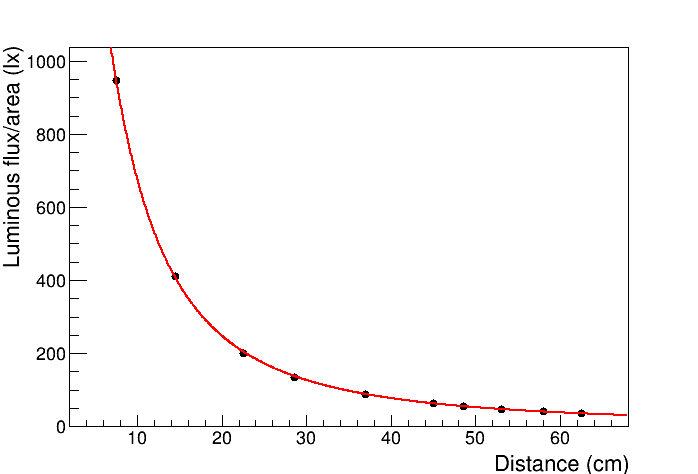}
\caption{Measured illuminance from lux meter as a function of distance between light source and the detector's plane.}
\label{fig:calgraph}
\end{figure}

Figure \ref{fig:calgraph} shows the obtained illuminance data as function of the distance $d$ from the closest point of the light source surface to the LDR. The circle markers show the data and the solid line shows the curve obtained with the parametrization of illuminance observed by the lux meter as a function of distance. A simple model was used for the curve least squares fit. It assumes the dimensional features of the source and sensors can be neglected and treated as punctual objects in good approximation and the illuminance is proportional to the inverse square of the source-sensor distance. The light angle of arrival on the lux meter surface is also taken into account in the model.

\begin{figure}[ht!]
\centering
\includegraphics[scale=0.4]{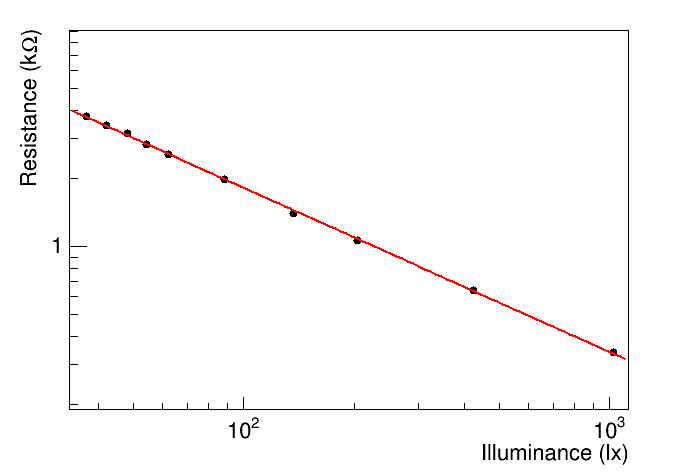}
\caption{Calibration of LDR resistance as function of illuminance.}
\label{fig:lnvsln}
\end{figure}

The obtained parametrization was used to estimate the illuminance arriving on the LDR sensor nearby. Figure \ref{fig:lnvsln} shows the 
measured resistance as a function 
of the illuminance estimated on the LDR. It is a very good approximation to assume the trend to be linear between these two quantities in a log-log scale. Therefore, data was linearized according to $\ln(R) = a + b \ln(I)$ where $R$ is the measured resistance and $I$ is the illuminance. Parameters $a$ and $b$ were extracted via a least squares fit providing a calibration between resistance and illuminance. Calibration is shown by the solid line with values obtained of $a = 3.95 \pm 0.02$ and $b = \rm -0.727\pm 0.005$. 

The calibration procedure adopted indicates that a LDR can be used as a measuring device provided the illuminances are in the range of $\rm 10-10^3$ lx. For a red wavelength source a $\rm \sim 5\%$ overall level of precision can be assumed due to calibration uncertainties only. 

We also reproduced the same procedures described in this section for other LDR sensors with $\rm 0.6\times0.7~cm^2$ and $\rm 0.9\times1.1~cm^2$ surfaces. A linear dependence on a log-log scale similar to the one seen in figure \ref{fig:calgraph} was found for all sensors. Table \ref{table:fit} summarizes the calibration constants obtained. Note the coefficients are all similar in value, despite sensor size differences, therefore, they should yield about the same precision for their measurements.

\begin{table}[ht!]
\caption{Calibration constants for the used LDR sensors, considering a dependence of the kind $\ln(R) = a + b \ln(I)$ .}
\label{table:fit}
\centering
\begin{tabular}{c|c|c}
\hline
Sensor Area $\rm(cm^2)$        & Coefficient $a$ & Coefficient $b$  \\ \hline
$\rm 0.4 \times 0.5$ & $3.95 \pm 0.02$    & $\rm -0.727\pm 0.005$ \\ \hline
$\rm 0.6 \times 0.7$ & $3.98 \pm 0.02$    & $\rm -0.755\pm 0.004$ \\ \hline
$\rm 0.9 \times 1.1$ & $3.58 \pm  0.02$   & $\rm -0.573\pm 0.004$ \\ \hline
\end{tabular}
\end{table}

This method offers an easier and cheaper way to establish a good quality calibration for the LDR than the procedure proposed in \cite{rbef} as it does not require a known laser light source and a set of adjustable polarizers to control irradiance. 

\section{Optics Experiments}\label{exps}

In this section we describe two simple optical experiments which were set up to use the LDR previously calibrated as sensor for illuminance measurements. The experimental results are presented in comparison with the theoretical expectations as a way to evaluate its feasibility as a light detector for didactic lab activities or simple demonstrations.

A red laser available in the laboratory was used as light source for both experiments proposed in the next sections. These are very easy and cheap to purchase and should not be a limiting factor for anyone wanting to reproduce any of the steps of this paper or to build their own experiments based on this work. Regarding the laser operation voltage one must make sure the appropriate power is supplied as fluctuations can alter the source power. Hence, one must either use charged up batteries or a voltage supply device which if not available can be improvised with an old mobile charger attached to a voltage divider to adapt tension. In order to reduce illuminance to values within calibration range ($\rm <10^3~lx$) we placed a 1mm plastic pinhole right in front of the light beam. 

Regarding the measurements of specular light directly incident on the sensor it was noticed that the $\rm 0.4 \times 0.5 ~ cm^2$ sensor provided the most stable values due to its higher density of semiconductor lines per sensor surface such that no variation was observed if the beam spot slightly moved over said area. Another detail noticed was that the positioning of the sensor has to be such that light must arrive at the sensor surface with an incoming angle as close to $\rm 90^o$ as possible. Given these observations, all the experiments were performed with the same sensor as these were not limiting factors to the measurements procedures. 

\subsection{Diffuse reflections on plaster}\label{exp0}

\begin{figure}[ht!]
\centering
\includegraphics[scale=0.5]{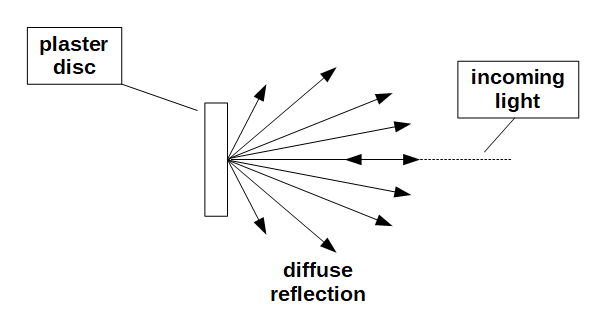}
\caption{Illustration of diffuse reflection of light on a plaster disc.}
\label{fig:plaster}
\end{figure}

A plaster disc was used as a target for the measurement of the typical diffuse reflection distribution exhibited by this material. Figure \ref{fig:plaster} shows an incoming light beam that reaches the plaster and is reflected diffusely in all directions as indicated by the set of outgoing arrows. The length of the arrows illustrate the expected illuminance as function of the angle of reflection with respect to the direction of a unitary vector perpendicular to the disc surface.

For an ideal diffusely reflecting material the radiance must be the same observed from any angle $\theta$ with respect to the normal to the surface which implies that the illuminance measurements obtained with the LDR can be modeled as: 
\begin{equation}
I(\theta) = \xi \cos(\theta) + \kappa,
\end{equation}
where $\xi$ relates to the incoming beam illuminance and $\kappa$ is a constant accounting for background. 

A certain amount of the background comes from external light impinging on the sensor as the experiment was performed with dimmed light only and not total darkness as for the calibration procedure in section \ref{calib}. The main reason for that choice was to verify reproducibility in a non-equipped experimental environment. It also turned up that the positioning and alignment of the sensor for each angle and its data reading is much better made with some visibility in the room.

\begin{figure}[ht!]
\centering
\includegraphics[scale=0.4]{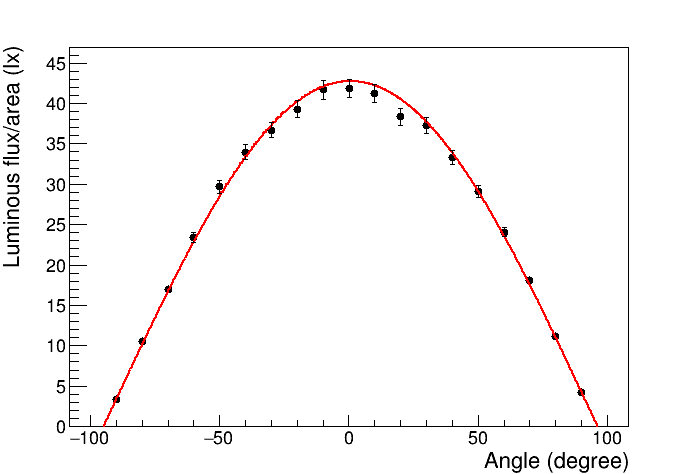}
\caption{Illuminance measured as function of scattering angle for laser beam shine onto a plaster disc.}
\label{fig:lambertian}
\end{figure}

The graph shown in figure \ref{fig:lambertian} was obtained by measuring the illuminance placing the LDR sensor at a fixed distance to the point of light reflection on the disk surface at different angles. The disk and laser source were kept at fixed positions and only the sensor was repositioned. The solid line illustrates the model fitted to the data. The quality of the data acquired with our proposed sensor allows to verify that the plaster material in fact presents a reflective behavior compatible mostly with a Lambertian reflector.   

It was noticed that for better performance of this experiment it was adequate to make a simple angular alignment procedure of the used protractor. This was done by comparing the measured illuminances of at least two equivalent fixed angles, $\theta$ and -$\theta$, checking if they gave about the same LDR readings. Random biases to data acquisition can be introduced as in this experiment the positioning and orientation of the sensor can be more challenging. Therefore, it is also good practice to take a few independent scan measurements and average the data sets for each angle in order to get more accurate estimates. Figure \ref{fig:lambertian} was obtained averaging two data sets.

\subsection{Transmissions and reflections on a glass slab}\label{exp1}

A glass slab was used as traversing media for the laser beam described in section \ref{exps} and the LDR sensor was used to measure the externally transmitted and reflected light as function of the beam incident angle. Because the slab is immersed in air, multiple internal reflections occur giving rise to multiple less bright parallel beams on both sides of the slab.

\begin{figure}[ht!]
\centering
\includegraphics[scale=0.5]{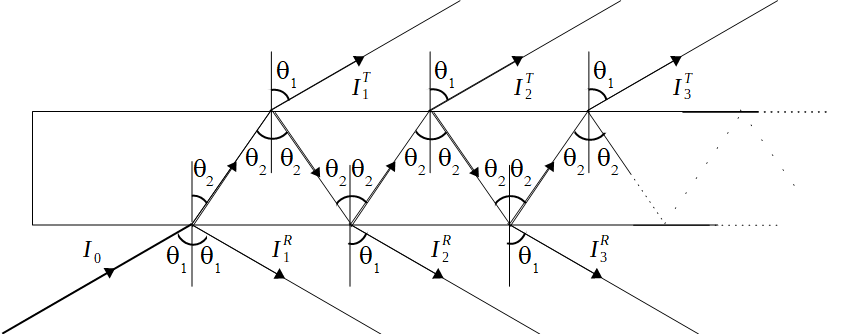}
\caption{Transmission and reflection profile of a glass slab with an incident light beam at a fixed angle $\theta_1$.}
\label{fig:glassslab}
\end{figure}

Figure \ref{fig:glassslab} illustrates a schematics of the experimental setup for a fixed impinging angle of the light beam. The incident angles and the transmissions and reflection points are indicated. Illuminance for each outgoing beam is represented by $I_i^{T,R}$ where $i$ is the number of the transmitted ($T$) or reflected ($R$) light beam.

The dimensions of the slab used were of $\rm 6~mm$ width, $\rm 2~cm$ height and $\rm 24.5~cm$ length. A scan on the beam angle was performed and the illuminance was measured for the first four beams of light, two transmitted and two reflected. It is easy to visually locate the position of each beam by placing a paper sheet in front of both sides of the slab. The angle range for the measurements was somewhat restricted due to geometrical limitations related to the beam diameter, sensor surface size and its positioning. The main issues restricting measurements in certain angular configurations regarded low illuminance of beams and also geometrical separation between beams as this distance can be smaller than the sensor's diameter.

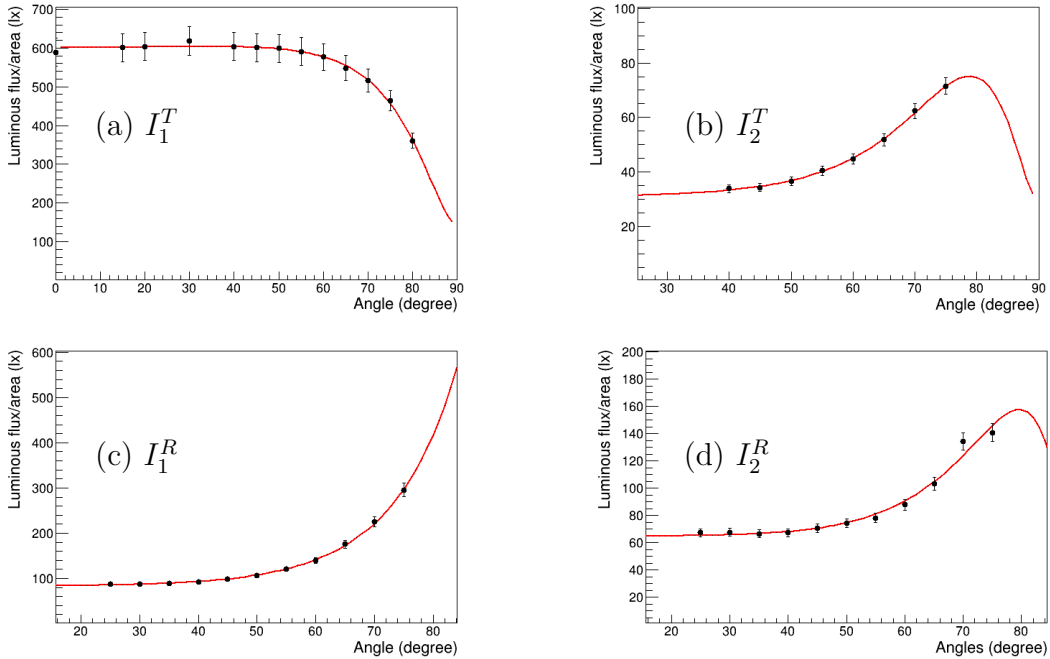
\begin{figure}
    \centering
    \begin{tikzpicture}
     \node[above right] (img) at (-0.5\textwidth,0) {\includegraphics[scale=0.27]{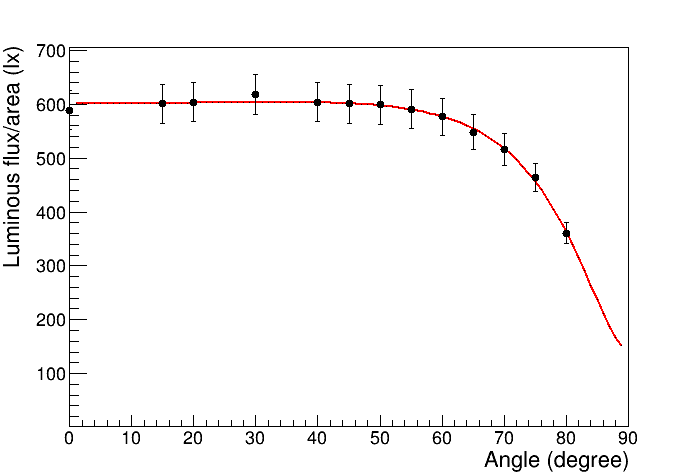}};
     \node at (-0.5\textwidth+0.12\textwidth,75pt) {(a) $I^T_1$};
     \node[above right] (img) at (-3pt,0pt)
     {\includegraphics[scale=0.27]{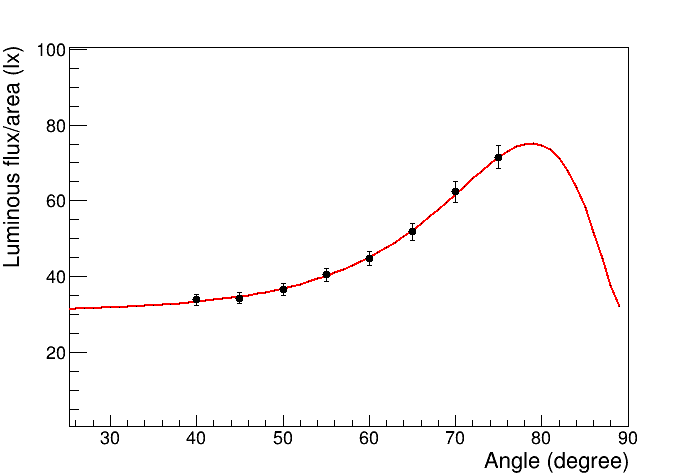}};
     \node at (0.12\textwidth,75pt) {(b) $I^T_2$}; 
     \node[above right] (img) at (-0.5\textwidth,-130pt)
    {\includegraphics[scale=0.27]{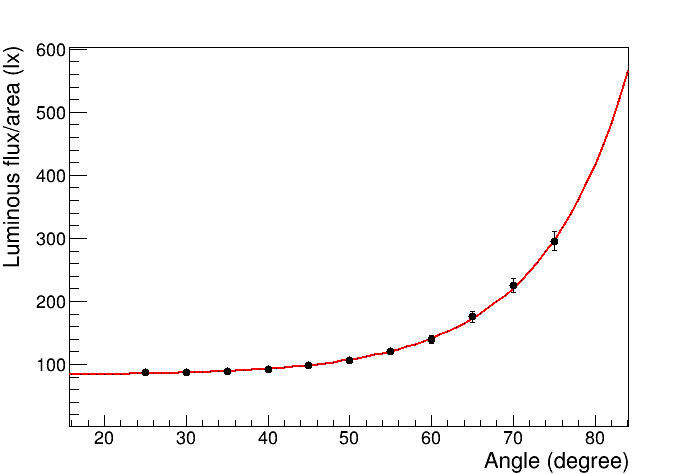}};
      \node at (-0.5\textwidth+0.12\textwidth,75pt-125pt) {(c) $I^{R}_1$};
      \node[above right] (img) at (0,-130pt)
    {\includegraphics[scale=0.27]{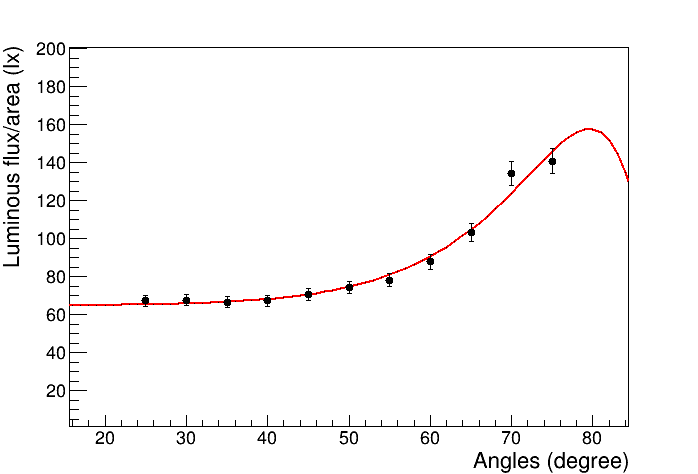}};
      \node at (0.12\textwidth,75pt-125pt) {(d) $I^{R}_2$}; 
     \end{tikzpicture}
    \caption{Transmitted and reflected illuminances for a laser beam incident on a glass slab as functions of $\theta$: (a) Transmitted beam 1, (b) Transmitted beam 2, (c) Reflected beam 1, (d) Reflected beam 2.}
    \label{fig:specular_t_r}
\end{figure}

The obtained illuminance as function of the angle for the first four beams are shown in figure \ref{fig:specular_t_r}. The points with error bars show the measured values as function of the angle of the incoming light beam from the source. The uncertainties bars are calculated considering the calibration uncertainties and also the estimated ambiguity from the alignment of the sensor with respect to the measured light beam ($\rm \sim10~\Omega$). This value was estimated as the spread of the readings obtained when placing the LDR in front of a given beam multiple times. The solid line in all graphs indicates the theoretical model curve fitted for each beam.  

The relative refractive index was obtained from each graph in figure \ref{fig:specular_t_r}. Table \ref{table:glassfit} shows the estimated values for the $\rm 650~nm$ wavelength. A satisfactory agreement among all estimates and the expected value (1.52) was found with percentage deviation below $\rm 5\%$. The estimated values were combined as a mean value, $n_{mean}$ and its uncertainty as $(n_{max}-n_{min})/2$.  

\begin{table}[ht!]
\caption{Estimated refractive index values for the glass slab to be compared with the expected value of 1.52 for glass.}
\label{table:glassfit}
\centering
\begin{tabular}{c|c|c|c|c}
\hline
$n_{I_1^T}$ & $n_{I_2^T}$ & $n_{I_1^R}$ & $n_{I_2^R}$ & $n_{mean}$ \\ \hline
1.51 & 1.50 & 1.58 & 1.45 & $1.51 \pm 0.06$ \\ \hline
\end{tabular}
\end{table}

The transmission ($t$) and reflection ($r$) coefficients for an uniform planar interface between two media where light propagates are given by \cite{griffiths}:
\begin{eqnarray}
\fl r_{TE} = \frac{E_r^{\perp}}{E_i^{\perp}}=\frac{\cos(\theta)-\sqrt{n^2-\sin^2(\theta)}}{\cos(\theta)+\sqrt{n^2-\sin^2(\theta)}}, &r_{TM} =\frac{E_r^{\parallel}}{E_i^{\parallel}}=\frac{2\cos(\theta)}{\cos(\theta)+\sqrt{n^2-\sin^2(\theta)}},\nonumber\\
\fl t_{TE} = \frac{E_t^{\perp}}{E_i^{\perp}}=\frac{-n^2\cos(\theta)+\sqrt{n^2-\sin^2(\theta)}}{n^2\cos(\theta)+\sqrt{n^2-\sin^2(\theta)}}, \quad &t_{TM}=\frac{E_t^{\parallel}}{E_i^{\parallel}}=\frac{2n\cos(\theta)}{n^2\cos(\theta)+\sqrt{n^2-\sin^2(\theta)}},
\end{eqnarray}
where the subscripts $TE$ and $TM$ indicate the cases for light polarization normal ($\perp$) and parallel ($\parallel$) to the incidence plane, $E_i$, $E_r$, $E_t$ are the amplitudes of the electric field of the incoming, reflected and transmitted light waves, $\theta$ is the light impinging angle and $n = n_{glass}/n_{air}$ is the relative refractive index. 

The two polarization expected contributions to transmittance ($\mathcal{T}$) and reflectance ($\mathcal{R}$) for each outgoing light beam on figure \ref{fig:glassslab} can be calculated by taking the magnitude square of these coefficients and multiplying them adequately. For instance, the reflectance expressions for the first reflected light beam ($i$=1) are given by $\mathcal{R} = |r|^2$ where $|r|$ is the reflection coefficient for light incoming from air onto the glass surface. The transmittance expressions for the first transmitted light beam are given by the product of the transmittance for light crossing from air to glass and the transmittance for light passing from glass to air. The same procedure is performed for the other light beams observed. Because the laser is a monocromatic source the expected illuminance for each beam, as expressed in equation \ref{ill_vs_irr}, can be modelled as:

\begin{equation}
I^T_i(\theta) = \alpha \, \mathcal{T}_{TM} + \beta\, \mathcal{T}_{TE} + \gamma \quad {\rm or} \quad I^R_i(\theta) = \delta\, \mathcal{R}_{TM} + \epsilon\, \mathcal{R}_{TE} + \zeta,
\end{equation}
where $i$ refers to the beam index, $\alpha,\beta, ...,\zeta$ are positive real numbers which account for the amount of light for the two polarizations in the incoming beam and also for a background baseline used for the same purpose as for the theoretical model presented on section \ref{exp0}.

However, another contribution for the background can be the result of the residual slightly off-beam light that can bounce internally many times and later comes off the finite glass slab in different outgoing angles reaching the sensor therefore adding to the constants values included in the fit. 

The data trends on graphs (a) and (c) of figure \ref{fig:specular_t_r} show decreasing and increasing only behaviors, respectively. Theoretically this follows from the fact that the transmittance and reflectance terms of the first two beams ($I^T_1$, $I^R_1$) are given by pure powers of $|t|$ or $|r|$ which have the same type of trends. On the other hand the following light beams ($I^T_i$, $I^R_i$ with $i>1$) do present a behavior that is similar to the graphs (b) and (d) where the terms are given by products between powers of $|r|$ and $|t|$ because of the internal reflections and consequent transmissions resulting on a slow rise of the illuminance as $\theta$ increases followed by a sharper decrease near the $\theta=90^o$ angle. 

Overall, the measurements obtained with the LDR for this simple glass slab setup provided data with good enough quality such that it is possible to make theoretical interpretations of the meaningful physics aspects of the experiment. 

\section{Discussion}

The understanding of optical processes is deeply related to many advances in science and technology as it has also directly contributed to many improvements in our day-to-day activities. For instance, the microscope and telescope are the most known iconic examples. But other important more contemporary developments of optical devices do employ these physical principles such as optical fibers, laser sources and semiconductor devices used in telecommunications, energy generation and medical diagnostics and therapy treatments. It is, therefore, a relevant and present topic that is applied in different areas of knowledge.

The use of hands-on experiments together with other resources such as applets, animated images, and simulations can be much beneficial to learning \cite{mit}, specially when a subject is rather abstract and highly mathematized, as multiple reflections and transmissions on interfaces between different media can be. Letting students become familiarized with the characteristics of a measuring device and then proceeding with a more qualitative and subsequent quantitative analysis of a phenomenon and discussions on the results has the potential to improve their understanding on a given subject, promoting a more active attitude towards learning. For students these kind of activities are rich opportunities which facilitate a more experiment based study similar to the actual work performed in physics research laboratories.

On the other hand, the reproduction in laboratory classes of the calibration and experiments presented by this paper can be performed in different ways depending on the didactic purpose intended and time available. In general, the acquisition of the data presented in sections \ref{calib} and \ref{exps} are relatively short and analyses can be made with the help of a computer and spreadsheet to speed up and avoid repetitive calculations. Although simple, performing the proposed experiments require adequate positioning and alignment of the materials so they are not displaced or rotated mid measurement scans. In some activities it might be of interest the students be left to figure out their own methods to set up an experiment. On other situations initial instructions can be given on how to run the experiment if a well defined procedure is expected to be done. For instance, in case the observation of the illuminance in a given situation is the only goal, the calibration step can be skipped if the experiment preparation is secondary. 

Considering the sensors small size and easy handling it should be possible to make other experiments similar to the ones presented in this paper. For example, one can use a semi circular perspex block to study the transmitted light beam illuminance angle dependence and determine the critical angle from the obtained curve. Another interesting experiment is to measure the illuminance of a TM polarized light beam  reflected on a glass slab as function of the incidence angle of the source beam. From the obtained curve it should be possible to extract the Brewster's angle as in \cite{benenson}. 

\section{Conclusions}\label{concl}

The feasibility of the use of LDR as suitable detectors for optical laboratory activities was evaluated. Calibration of LDRs with satisfactory precision ($\rm \sim 5\%$) was obtained with a simple procedure using low-cost materials such as a LED red lamp and a lux meter. Minimalistic optical experiments were also performed as proof of principle for illuminance measurements giving good quality estimates for multiple transmitted and reflected light beams as function of incidence angle of an incoming red laser beam onto a glass slab and also for the angular distribution of diffusely reflected light from a plaster disc. In both experiments data allowed to confirm angular dependencies in a very clear manner in the form of graphs without excessive data analysis procedures. The extraction of the refraction index also provided estimates in good agreement with the expected value. 

Hence, the LDRs are versatile sensors that can in principle be used in a range of optics experiments where measurements of diffuse and specular light reflection are required. They are easy to use and their required data acquisition time intervals are adequate for didactic activities. The determination of material optical properties from the illuminance dependence with angles is accessible, therefore, providing possibilities for a broader understanding of the physics phenomena via an experimental approach.

\ack{F. Marinho and L. Paulucci would like to thank Funda\c c\~ao de Amparo \`a Pesquisa do Estado de S\~ao Paulo (FAPESP) for financial support under grant $\rm n^o$  2017/13942-5.\\}

\bibliographystyle{unsrt}
\bibliography{references}
\end{document}